\documentclass[11pt]{article}
\usepackage{amsmath,amsthm}

\def\p{\partial}

\newtheorem*{teo}{Theorem}
\newtheorem*{cor}{Corollary}
\newtheorem{lem}{Lemma}

\newenvironment{dok}
{\noindent{\it Proof.}}{\hfill\par}

\newcommand{\la}{\lambda}

\newcommand{\ds}{\displaystyle}

\newcommand{\sca}[1]{\Bigl\langle #1\Bigr\rangle}

\begin{document}

\title{The Generating Formula for The Solutions to The Associativity Equations}
\author{A.\,A.\,Akhmetshin${}^*$,\qquad I.\,M.\,Krichever${}^*$,\qquad 
Y.\,S.\,Volvovski\thanks
{Columbia University, 2990 Broadway, New York, NY 10027, USA and 
Landau Institute for Theoretical Physics, Kosygina 2, 117940 Moscow, Russia}}

\date{March 9, 1999}

\maketitle

\begin{abstract}
An exact formula for the solutions to the WDVV equation in terms of horizontal
sections of the corresponding flat connection is found.
\end{abstract}

\noindent{\bf 1.~Introduction.}
The associativity equations (or WDDV equations) appeared in the classification 
problem for the  topological field theories at the early 90's, 
see \cite{dvv}, \cite{dw}.
During the last years these equations has attracted a great interest due 
to connections with the enumerative geometry
(Gromov\,--\,Witten invariants \cite{km}), quantum cohomology \cite{giv}
and the Whitham theory \cite{kr1}.

In \cite{dub} it was noticed that the classification problem for
the topological field theories is equivalent to the classification of the 
Egoroff metrics of special type. Egoroff metrics are flat diagonal metrics 
\begin{equation}\label{ds}
ds^2=\sum\limits_{i=1}^n h^2_i(u) (du^i)^2,\qquad u=(u^1,\dots,u^n)
\end{equation}
such that $\p_i h^2_j(u)=\p_j h^2_i(u)$, where $\p_i=\p/\p u^i$.  
It turns out that for every Egoroff metric with the additional constraints
$\sum_{j=1}^n \p_j h_i=0$  the functions
\begin{equation}\label{c1}
c_{kl}^m(x) = \sum_{i=1}^n \frac{\p x^m}{\p u^i} \frac{\p u^i}{\p x^k} 
\frac{\p u^i}{\p x^l},
\end{equation}
where $x^k(u)$ are the flat coordinates of the metric (\ref{ds}), satisfy
\begin{equation}\label{c2}
c_{ij}^k(x) c_{km}^l(x) = c_{jm}^k(x) c_{ik}^l(x) ,
\end{equation}
which are the associativity conditions for the algebra 
$\phi_k\phi_l=c_{kl}^m\phi_m$.
Moreover, there exists a function $F(x)$ such that its third derivatives are 
equal to
\begin{equation}\label{f1}
\frac{\p^3 F(x)}{\p x^k \p x^l \p x^m}=c_{klm}(x)=\eta_{mi} c_{kl}^i(x),\quad
\mbox{where}\quad \eta_{pq}=\sum_{i=1}^n h_i^2(u) \frac{\p u^i}{\p x^p} 
\frac{\p u^i}{\p x^q} .
\end{equation}
In the topological field theory with $n$ primary fields $\phi_k$
function $F$ plays role of the partition function.
There also exist constants $r^m$ such that $\eta_{kl}=r^m c_{klm}(x)$.

Equations (\ref{c2}) and the existence of the function $F$ such that 
(\ref{f1}) holds are equivalent to the compatibility conditions of the 
linear system (see \cite{dub})
\begin{equation}\label{ph}
\p_k \Phi_l -\lambda c_{kl}^m \Phi_m =0 ,
\end{equation}
where $\lambda$ is a spectral parameter.

In few special cases the function $F$ was found explicitly 
(see \cite{dub}, \cite{kr3}, \cite{dpk}).
However, in general case an expression for $F$ in terms of the flat sections 
of the connection $\nabla_k=\p/\p x^k-\la c_{kl}^m$ has been unknown. 
The main goal of these notes is to write down such an expression.
It was motivated by the results of \cite{kr2}, where explicit formulas 
for the algebraic-geometrical solutions to the associativity equations were
found. Though the general formula of \cite{kr2} (Theorem~5.1) is correct, 
in the final formula (Theorem~5.2) one of the terms was missed. 
Here we take an opportunity to fix it.

\bigskip

\noindent{\bf 2.}
Let us consider a solution $\beta_{ij}(u)=\beta_{ji}(u)$ to the 
Darboux\,--\,Egoroff system
\begin{equation}\label{la}
\p_k\beta_{ij}=\beta_{ik}\beta_{kj},\quad i\ne j\ne k;\qquad\quad 
\sum_{m=1}^n \p_m\beta_{ij}=0, \quad i\ne j.               
\end{equation}
Following \cite{dub} we fix the unique Egoroff metric (\ref{ds})
corresponding to this solution by defining the Lame coefficients~$h_i(u)$
from the system  
\begin{equation}\label{h}
\p_j h_i(u)=\beta_{ij}(u) h_j(u),\quad i\ne j,\qquad\quad     
\p_i h_i(u)=-\sum_{j\ne i}\beta_{ij}(u) h_j(u),     
\end{equation}
and initial conditions $h_i(0)=1$, $i=1,\dots,n$.
This system is compatible due to (\ref{la}).
The condition $\beta_{ij}=\beta_{ji}$ implies 
that the above defined metric is in fact Egoroff metric.

The flat coordinates $x^1,\dots,x^n$ of this metric can be found from  
the linear system
\begin{equation}\label{x}
\p_i\p_j x^k=\Gamma_{ij}^i \p_i x^k + \Gamma_{ji}^j \p_j x^k ,
\quad i\ne j;\qquad\quad                                                 
\p_i\p_i x^k=\sum_{j=1}^n \Gamma_{ii}^j \p_j x^k ,          
\end{equation}
where $\Gamma_{ij}^k$ are Christoffel symbols: 
$\Gamma_{ij}^i=\p_j h_i/h_i$, 
$\Gamma_{ii}^j=(2\delta_{ij}-1)(h_i\p_j h_i)/(h^2_j)$. 
We choose the following initial conditions: $x^k(0)=0$,
$\sum_{k,l} \eta_{kl} \,\p_i x^k(0) \p_j x^l(0)=\delta_{ij}$.
Here $\eta_{kl}$ is the fixed symmetric nondegenerate matrix.

The Darboux\,--\,Egoroff system can be regarded as the compatibility 
conditions of the following linear system: 
\begin{align}
\p_j \Psi_i(u,\la) &=\beta_{ij}(u) \Psi_j(u,\la), \qquad i\ne j \label{ps1}\\
\p_i \Psi_i(u,\la) &=\la\Psi_i(u,\la)-\sum_{k\ne i}\beta_{ik}(u)\Psi_k(u,\la),
                                                                \label{ps2} 
\end{align} 
Here $\Psi_i=(\Psi_i^1,\dots,\Psi_i^n)$, $i=1,\dots,n$, are vector-functions 
which are formal power series in $\la$. There exists a unique solution of this 
linear system with the initial conditions $\Psi^k_i(0,\la)=\la \p_i x^k(0)$.

It follows from (\ref{ps1}), (\ref{ps2}) that the expansion of 
$\Psi_i$ has the form
$$
\Psi_i(u,\la)=\sum_{s=0}^{\infty} \frac{\p_i\xi_s^k(u)}{h_i(u)} \la^s,
$$
where $\xi_0^k=r^k$ are constants (we will define them later), 
$\xi_1^k(u)=x^k(u)$ are flat coordinates and $\xi_s^k$ for $s\ge 2$ obey 
the following equations
\begin{align}
\p_i\p_j \xi_s^k &=\Gamma_{ij}^i \p_i \xi_s^k + \Gamma_{ji}^j \p_j \xi_s^k ,
\qquad i\ne j,                                           \label{xi1}\\
\p_i\p_i \xi_s^k &=\sum_{j=1}^n \Gamma_{ii}^j \p_j \xi_s^k +\p_i\xi_{s-1}^k , 
                                                         \label{xi2}  
\end{align}
with the initial conditions $\xi_s^k(0)=0$, $\p_i\xi_s^k(0)=0$.
The equations
\begin{equation}\label{sec}
\frac{\p^2 \xi_s^m}{\p x^k\p x^l}=\sum_{p=1}^nc_{kl}^p 
\frac{\p \xi_{s-1}^m}{\p x^p},
\end{equation}
where $c_{kl}^p$ are defined by (\ref{c1}), follow immediately 
from (\ref{xi1}), (\ref{xi2}). We will denote $\xi_2^k(u)$ and $\xi_3^k(u)$ 
by $y^k(u)$ and $z^k(u)$, respectively. 

System (\ref{ps1},\ref{ps2}) implies that $\la\Psi_i=\sum_{j=1}^n \p_j\Psi_i$ 
and, therefore, $\sum_{i=1}^n \p_i \xi_s^k(u)=\xi_{s-1}^k$ for $s\ge 1$. 
The last equality for $s=1$ is, in fact, the definition of the constants $r^k$.

\bigskip

\noindent{\bf 3.}
Let $\psi$ be given by the formula
\begin{equation}\label{psi}
\la\psi(u,\la)=\sum_{i=1}^n h_i(u)\Psi_i(u,\la) .
\end{equation}
It's straightforward to check that $\p_i\psi(u,\la)=h_i(u)\Psi_i(u,\la)$.
As a formal power series on~$\la$ the $k$-th component of the vector-function
$\psi(u,\la)$ has the form
\begin{equation}
\psi^k(u,\la)=r^k+x^k(u)\la+y^k(u)\la^2+z^k(u)\la^3+
\sum_{s=4}^{\infty} \xi_s^k(u)\la^s.
\end{equation}
Note that (\ref{sec}) implies that the functions $\Phi_k(x)=\p \psi(x)/\p x^k$ 
satisfy (\ref{ph}). Moreover,
$$
\la\psi(x)=\sum_{k=1}^n r^k\Phi_k(x)
$$ 
and thus the function $\psi$ can be regarded as a generating function for
the  flat sections of the above-defined connection $\nabla_k$.

\begin{lem}
The vector-functions $\Psi_i(u,\la)$ satisfy the equations
\begin{equation}\label{or}
\sca{\Psi_i(u,\la),\Psi_j(u,-\la)}=-\delta_{ij}\la^2 , 
\end{equation}
the scalar product given by the matrix $\eta_{kl}$: 
$\langle A,B\rangle=\eta_{kl} A^k B^l$.
\end{lem}

\begin{lem}
The scalar product 
\begin{equation}\label{or2}
\sca{\frac{1}{\la}\p_i\psi(u,\la),\psi(u,-\la)}
\end{equation}
does not depend on $\la$.
\end{lem}

Both lemmas can be proved in the similar way. First we establish the
required properties of the scalar products (\ref{or}), (\ref{or2}) at 
the point $u=0$. Then we show that they satisfy certain differential 
equations. The Uniqueness Theorem implies both statements. 

\begin{cor}
The functions $x^k(u)$, $y^k(u)$ and $z^k(u)$ satisfy the following
relation:
\begin{equation}\label{xyz}
\sum_{q=1}^n \eta_{kq} y^q = \sum_{p,q=1}^n \eta_{pq}\left(
x^q\frac{\p y^p}{\p x^k}-r^q\frac{\p z^p}{\p x^k} \right) .
\end{equation}
\end{cor}

\begin{dok}
Consider the scalar product
$\ds\sca{\frac{1}{\la}\frac{\p\psi}{\p x^k}(u,\la),\psi(u,-\la)}$.
Since $\p\psi/\p x^k$ is the linear combination of $\p\psi/\p u^i$ Lemma~2  
implies that this scalar product is independent of~$\la$.
On the other hand, it can be presented as a power series in $\la$
\begin{equation}\label{sp2}
\sca{\frac{1}{\la}\frac{\p\psi}{\p x^k},\psi^{\sigma}}=
\frac{1}{\la}\sum_{p,q=1}^n \eta_{pq}\left( \delta_k^p\la + 
\frac{\p y^p}{\p x^k}\la^2+\frac{\p z^p}{\p x^k}\la^3+\ldots \right) 
\left( r^q-x^q\la+y^q\la^2-z^q\la^3+\ldots \right) 
\end{equation} 
(here $\psi^{\sigma}(u,\la)=\psi(u,-\la)$).
Therefore, all but the first coefficients of the series (\ref{sp2}) should
equal zero. Applying this argument to the coefficient of~$\la^2$ 
we obtain~(\ref{xyz}). 
\end{dok}

\begin{teo}
The function $F(x)=F(u(x))$ defined by the formula 
\begin{equation}
F(u)=\frac12  \sum_{p,q=1}^n \eta_{pq} \left(x^q(u) y^p(u)-r^q z^p(u)\right)
\end{equation}
satisfies the equation (\ref{f1}).
\end{teo}

\begin{dok}
Let us notice that Corollary implies $\p F/\p x^k=\sum_{q=1}^n \eta_{kq} y^q$. 
Now the statement of the Theorem is direct implication of (\ref{c1}) and
(\ref{sec}) for~$s=2$.
\end{dok}

From the previous formulas it follows that $F$ satisfies the renormalization 
group type equation:
$$
F(x)-\sum_{k=1}^n x^k\frac{\p F}{\p x^k}=-\sum_{p,q=1}^n \eta_{pq} r^q z^p.
$$ 
In our next paper we hope to obtain more general formula, where the 
function $F$ depends on infinitely many variables corresponding to 
gravitational descendants of the primary fields~$\phi_k$.

\end{document}